\documentstyle[12pt,fleqn]{article}

\font\twlgot =eufm10 scaled \magstep1
\font\egtgot =eufm8
\font\sevgot =eufm7
\font\twlmsb =msbm10 scaled \magstep1
\font\egtmsb =msbm8
\font\sevmsb =msbm7

\newfam\gotfam
\def\pgot{\fam\gotfam\twlgot}
\textfont\gotfam\twlgot
\scriptfont\gotfam\egtgot
\scriptscriptfont\gotfam\sevgot
\def\got{\protect\pgot}
\newfam\msbfam
\textfont\msbfam\twlmsb
\scriptfont\msbfam\egtmsb
\scriptscriptfont\msbfam\sevmsb
\def\Bbb{\protect\pBbb}
\def\pBbb{\relax\ifmmode\expandafter\Bb\else\typeout{You cann't use
Bbb in text mode}\fi}
\def\Bb #1{{\fam\msbfam\relax#1}}

\newcommand{\gO}{{\got O}}
\newcommand{\gQ}{{\got T}}

\newcommand{\gE}{{\got E}}
\newcommand{\gA}{{\got A}}

\newcommand{\gd}{{\got d}}

\newcommand{\gS}{{\got S}}

\def\thebibliography#1{\section*{References}\list
  {[\arabic{enumi}]}{\settowidth\labelwidth{#1}\leftmargin\labelwidth
    \advance\leftmargin\labelsep
    \usecounter{enumi}}
    \def\newblock{\hskip .11em plus .33em minus .07em}
    \sloppy\clubpenalty4000\widowpenalty4000
    \sfcode`\.=1000\relax}

\def\op#1{\mathop{\fam0 #1}\limits}

\newcommand{\nm}[1]{|{#1}|}
\newcommand{\beq}{\begin{equation}}
\newcommand{\eeq}{\end{equation}}
\newcommand{\ben}{\begin{eqnarray}}
\newcommand{\een}{\end{eqnarray}}
\newcommand{\be}{\begin{eqnarray*}}
\newcommand{\ee}{\end{eqnarray*}}
\newcommand{\bea}{\begin{eqalph}}
\newcommand{\eea}{\end{eqalph}}
\newcommand{\cA}{{\cal A}}

\newcommand{\cP}{{\cal P}}

\newcommand{\cL}{{\cal L}}

\newcommand{\cV}{{\cal V}}
\newcommand{\cE}{{\cal E}}

\newcommand{\cQ}{{\cal Q}}

\newcommand{\cS}{{\cal S}}
\newcommand{\cC}{{\cal C}}
\newcommand{\cO}{{\cal O}}
\newcommand{\bL}{{\bf L}}

\newcommand{\bs}{{\bf s}}

\newcommand{\al}{\alpha}
\newcommand{\vr}{\varrho}

\newcommand{\dl}{\delta}
\newcommand{\la}{\lambda}
\newcommand{\La}{\Lambda}
\newcommand{\f}{\phi}
\newcommand{\om}{\omega}

\newcommand{\m}{\mu}

\newcommand{\G}{\Gamma}
\newcommand{\th}{\theta}

\newcommand{\vt}{\vartheta}
\newcommand{\vf}{\varphi}
\newcommand{\up}{\upsilon}

\newcommand{\di}{{\rm dim\,}}

\newcommand{\si}{\sigma}
\newcommand{\Si}{\Sigma}
\newcommand{\w}{\wedge}

\newcommand{\ol}{\overline}
\newcommand{\dr}{\partial}
\newcommand{\ar}{\op\longrightarrow}
\newcommand{\ot}{\otimes}
\newcommand{\ap}{\approx}

\newcounter{eqalph}
\newcounter{equationa}
\newcounter{theorem}
\newcounter{remark}
\newcounter{proposition}
\newcounter{lemma}
\newcounter{corollary}
\newcounter{definition}

\newenvironment{eqalph}{\stepcounter{equation}
\setcounter{equationa}{\value{equation}}
\setcounter{equation}{0}

\begin{eqnarray}}{\end{eqnarray}\setcounter{equation}{\value{equationa}}}

\def\theremark{\arabic{remark}}
\def\thetheorem{\arabic{theorem}}

\newenvironment{proof}{\noindent 
{\it Proof.}}{\hfill $\Box$ \medskip}

\newenvironment{theo}{\refstepcounter{theorem} 
\bigskip\noindent{\sc THEOREM \thetheorem.} \it}{\medskip}
\newenvironment{prop}{\refstepcounter{theorem} 
\bigskip\noindent{\sc PROPOSITION \thetheorem.}\it}{\medskip}

\newenvironment{cor}{\refstepcounter{theorem} 
\bigskip\noindent{\sc COROLLARY \thetheorem.}\it}{\medskip}

\newcommand{\mar}[1]{}

\begin{document}
\hbox{}

{\parindent=0pt

{\large\bf Lagrangian symmetries and supersymmetries 
depending on derivatives. Conservation laws and cohomology}
\bigskip

{\sc G. GIACHETTA}$^1$, {\sc L. MANGIAROTTI}$^2$ and {\sc G.SARDANASHVILY}$^3$
\bigskip

\begin{small}

$^1${\it Department of Mathematics and Informatics, University of Camerino, 62032
Camerino (MC), Italy. e-mail: giovanni.giachetta@unicam.it}

$^2${\it Department of Mathematics and Informatics, University
of Camerino, 62032 Camerino (MC), Italy. e-mail:
luigi.mangiarotti@unicam.it}

$^3${\it Department of Theoretical Physics, Physics Faculty,
Moscow State University, 117234 Moscow, Russia. e-mail
sard@grav.phys.msu.su}
\bigskip

{\bf Abstract.} 
Motivated by BRST theory, we study 
generalized symmetries and
supersymmetries depending on derivatives of dynamic variables
in a most general setting.
We state the first variational formula and conservation laws 
for higher order Lagrangian systems on fiber
bundles and graded manifolds under  generalized symmetries and
supersymmetries of any order.
Cohomology of nilpotent generalized supersymmetries
are considered. 

\medskip

\medskip


\end{small}
}

\section{Introduction}

Symmetries of differential equations under transformations
of dynamic variables depending on their derivatives 
have been intensively investigated (see
\cite{and93,kras,olv} for a survey). 
Following \cite{and93,olv}, we agree to call them the
generalized symmetries in contrast with classical (point)
symmetries. In mechanics, conservation laws corresponding to
generalized symmetries are well known \cite{olv}. In field
theory, BRST transformations provide the most 
interesting example of generalized symmetries
\cite{fat,fulp},
 but they involve odd
ghost fields. Therefore, we aim to consider both 
generalized symmetries of classical Lagrangian systems on smooth
fiber bundles and generalized supersymmetries of Lagrangian
systems on graded manifolds.
 
Generalized symmetries of Lagrangian
systems on a local
coordinate domain of a trivial fiber bundle have been described
in detail \cite{olv}. In the recent work
\cite{fat}, a global analysis of first order
Lagrangian systems and conservation laws under generalized
symmetries depending on first order derivatives has been
provided. We aim studying the conservation laws
in higher order Lagrangian systems on fiber bundles and
graded manifolds under generalized symmetries and
supersymmetries of any order.
Let us emphasize the following.

(i) An $r$-order differential equation on a fiber bundle
$Y\to X$ is conventionally defined as a closed subbundle of the
$r$-order jet bundle $J^rY\to X$
of sections of $Y\to X$ \cite{bry,kras}.
Euler--Lagrange equations need not satisfy this condition,
unless an Euler--Lagrange operator is of constant rank.
Therefore, we regard infinitesimal symmetry transformations of
Lagrangians and Euler--Lagrange operators as
differential operators on a graded differential algebra
(henceforth GDA) of exterior forms, but not as manifold maps. 
For instance,  we are not concerned with dynamic symmetries.
This approach is straightforwardly extended to Lagrangian
systems on graded manifolds. 

(ii) We use the first variational
formula in order to obtain Lagrangian conservation laws.
Recall that an $r$-order Lagrangian of a Lagrangian system on a
fiber bundle $Y\to X$ is defined as a horizontal density
$L:J^rY\to
\op\w^nT^*X$,
$n=\di X$, on the $r$-order jet manifold $J^rY$. Let $u$ 
be a
projectable vector field on $Y\to X$ seen as an
infinitesimal generator of
a local one-parameter group of bundle automorphisms 
of $Y\to X$. Let $\bL_{J^ru}L$ be the Lie derivative of
$L$ along the jet prolongation $J^ru$ of $u$ onto $J^rY$.
The first
variational formula provides its canonical decomposition 
\mar{g16}\beq
\bL_{J^ru}L=u_V\rfloor\dl L + d_H(h_0(J^{2r-1}u\rfloor\Xi_L)), 
\label{g16}
\eeq
where $\dl L$ is the Euler--Lagrange operator, $\Xi_L$ is
a Lepagean equivalent of $L$ (e.g., a Poincar\'e--Cartan form),
$u_V$ is a vertical part of
$u$,
$d_H$ is the total differential, and $h_0$ is the
horizontal projection (see their definitions below)
\cite{book,book00,epr}.  Let $u$ be a
divergence symmetry of $L$, i.e., the Lie
derivative 
$\bL_{J^ru}L$ is a total differential $d_H\si$.
Then, the first variational
formula (\ref{g16}) on the  kernel of
the Euler--Lagrange operator $\dl L$ leads to
the conservation law
\mar{g19}\beq
0\ap d_H(h_0(J^{2r-1}u\rfloor\Xi_L)-\si). \label{g19}
\eeq
If $u$ is a (variational) symmetry  of $L$, i.e.,
$\bL_{J^ru}L=0$, the conservation law (\ref{g19})
comes to the familiar Noether one. Our goal is to extend the
first variational formula (\ref{g16}) to generalized symmetries 
and supersymmetries, and to obtain the corresponding
Lagrangian conservation laws.

(iii) A vector field $u$ in the first variational formula 
(\ref{g16}) is a derivation of the $\Bbb R$-ring
$C^\infty(Y)$ of smooth real functions on $Y$. Let $\vt$ be a
derivation of 
$C^\infty(Y)$ with values in the ring $C^\infty(J^rY)$ of
smooth real functions on the jet manifold $J^kY$. It is called a
$k$-order generalized vector field. A generalized
symmetry can be defined as the prolongation $J^r\vt$ of $\vt$
onto  any finite order jet manifold $J^rY$.  This definition
recovers both the notion of a local generalized symmetry in
\cite{olv} and the definition of a generalized vector field
as a section of the pull-back bundle
$TY\times J^kY\to J^kY$ in \cite{fat}. 
The key point is that, in general, $\bL_{J^r\vt}\f$ is 
an exterior form on the jet manifold $J^{r+k}Y$. By virtue of
the well-known B\"acklund theorem,
$\bL_{J^r\vt}$ preserves the GDA 
$\cO^*_r$ of exterior forms on $J^rY$
iff either $\vt$ is a vector field on $Y$ or $Y\to X$ is a
one-dimensional bundle and $\vt$ is a generalized vector field
at most of first order. Thus, considering generalized
symmetries, we deal with Lagrangian systems of unspecified
finite order. Infinite order jet formalism 
\cite{ander,kras,book00,epr,tak2}  provides a
convenient tool for studying these systems both on fiber
bundles and graded manifolds. In the framework of this
formalism, the first variational formula issues from
the variational bicomplex, whose cohomology provides some
topological obstruction to generalized symmetries and
supersymmetries. For instance, if $\vt$ is a divergence
symmetry of a Lagrangian
$L$, the equality
\mar{g51}\beq
\dl (\bL_{J^r\vt}L)=0 \label{g51}
\eeq
holds, but the converse is not true. There is a topological
obstruction to $\vt$ in (\ref{g51}) to be a divergence
symmetry. At the same time,   one can think of the equality
(\ref{g51}) as being at least locally the characteristic
equation for  divergence symmetries of a given Lagrangian
$L$. 
Recall that, by virtue of the master identity
\be
\bL_{J^{2r}u}\dl L=\dl(\bL_{J^ru}L), 
\ee  
any classical divergence symmetry of a
Lagrangian
is also a symmetry of its Euler--Lagrange operator.
However, this identity is not extended  to generalized
symmetries \cite{olv}. 

The following peculiarities of generalized
supersymmetries should be additionally noted. They are
expressed into jets of odd variables, and they can be nilpotent.

We do not concern particular geometric models of ghost
fields in gauge theory, but consider Lagrangian systems of odd
variables in a general setting. 
For this purpose, one calls into play fiber bundles over
graded manifolds or supermanifolds 
\cite{cari,cia,mont}. However, Lagrangian BRST
theory on $X=\Bbb
R^n$ \cite{barn,bran,bran01} involves jets of odd fields
only with respect to space-time coordinates. Therefore, we
describe odd variables on a smooth manifold
$X$ as generating elements of the structure ring of
a graded manifold whose body is
$X$. By the well-known Batchelor theorem \cite{bart},
any graded manifold is
isomorphic to the one whose structure sheaf is the sheaf
$\gA_Q$ of germs of sections of the exterior product 
\mar{g80}\beq
\w Q^*=\Bbb R\op\oplus_X
Q^*\op\oplus_X\op\w^2 Q^*\op\oplus_X\cdots,
\label{g80}
\eeq 
where $Q^*$ is the dual of some real vector
bundle $Q\to X$. In physical models, a vector bundle $Q$ is
usually given from the beginning. Therefore, we restrict our
consideration to so called simple graded manifolds
$(X,\gA_Q)$ where the
Batchelor isomorphism holds fixed. We agree to say that
$(X,\gA_Q)$ is constructed from $Q$. Accordingly,
$r$-jets of odd variables are defined as generating elements
of the structure ring of the simple graded manifold
$(X,\gA_{J^rQ})$ constructed from the jet bundle $J^rQ$ of
$Q$ \cite{book00,mpla}. This definition of jets 
differs from that of jets of a graded
fiber bundle in \cite{hern}, but reproduces 
the heuristic notion of jets of ghosts
in Lagrangian BRST theory on $\Bbb R^n$ in
\cite{barn,bran,bran01}.  Moreover, this definition enables  
one to study Lagrangian systems on a graded
manifold similarly to those on a fiber bundle.

The BRST transformation in gauge theory on a principal bundle
exemplifies a generalized supersymmetry (see $\up$
(\ref{g130}) below). The BRST operator is defined as the Lie
derivative $\bL_\up$ along this generalized symmetry. The
fact that it is nilpotent on horizontal (local in the
terminology of \cite{barn,bran}) forms motivates us to study
nilpotent generalized supersymmetries. They are necessarily
odd, i.e., there is no nilpotent generalized symmetry. The key
point is that the Lie derivative
$\bL_\up$ along a generalized supersymmetry and the total
differential
$d_H$ mutually commute. If $\bL_\up$ is nilpotent, we obtain a
bicomplex whose iterated cohomology
classifies Lagrangians with a given nilpotent divergence
symmetry.

\section{Lagrangian systems of unspecified finite order on
fiber bundles}

Finite order jet manifolds make up the 
inverse system
\mar{5.10}\beq
X\op\longleftarrow^\pi Y\op\longleftarrow^{\pi^1_0} J^1Y
\longleftarrow \cdots J^{r-1}Y \op\longleftarrow^{\pi^r_{r-1}}
J^rY\longleftarrow\cdots.
\label{5.10}
\eeq
Its projective limit $J^\infty Y$,
called the infinite order jet space,
is endowed with the weakest
topology such that surjections $\pi^\infty_r:J^\infty Y\to
J^rY$ are continuous. This
topology makes $J^\infty Y$ into a paracompact Fr\'echet
manifold \cite{tak2}. 
Any bundle coordinate atlas
$\{U_Y,(x^\la,y^i)\}$ of $Y\to X$ yields the manifold 
coordinate atlas 
\mar{jet1}\beq
\{(\pi^\infty_0)^{-1}(U_Y), (x^\la, y^i_\La)\}, \qquad
{y'}^i_{\la+\La}=\frac{\dr x^\m}{\dr x'^\la}d_\m y'^i_\La,
\qquad
0\leq|\La|,
\label{jet1}
\eeq
of $J^\infty Y$ where $\La=(\la_k...\la_1)$ is
a symmetric multi-index 
of length $k$, $\la+\La=(\la\la_k...\la_1)$, 
and  
\mar{5.177}\beq
d_\la = \dr_\la + \op\sum_{|\La|\geq 0}
y^i_{\la+\La}\dr_i^\La, \qquad
d_\La=d_{\la_k}\circ\cdots\circ d_{\la_1}, \quad
\La=(\la_k...\la_1), \label{5.177}
\eeq
are the total derivatives. Hereafter, we 
fix an atlas of $Y$ and, consequently, that of $J^\infty Y$
containing a finite number of charts, though their branches
$U_Y$ need not be domains \cite{greub}. 

With the inverse system
(\ref{5.10}), we have the
direct system 
\mar{5.7}\beq
\cO^*(X)\op\longrightarrow^{\pi^*} \cO^*(Y) 
\op\longrightarrow^{\pi^1_0{}^*} \cO_1^* \cdots
\op\longrightarrow^{\pi^r_{r-1}{}^*}
 \cO_r^* \longrightarrow\cdots \label{5.7}
\eeq
of the GDAs of exterior forms on
finite order jet manifolds with respect to 
the pull-back monomorphisms $\pi^r_{r-1}{}^*$. Its direct
limit
$\cO_\infty^*$ is a GDA, whose de Rham
cohomology equals that of the fiber bundle $Y$ \cite{ander}.
Though
$J^\infty Y$ is not a smooth manifold,  one can think of
elements of
$\cO_\infty^*$ as being objects on $J^\infty Y$ as
follows. 
Let $\gO^*_r$ be the sheaf
of germs of exterior forms on the $r$-order jet 
manifold $J^rY$, and let
$\ol\gO^*_r$ be its canonical presheaf.  There is the direct 
system of presheaves
\be
\ol\gO^*_X\op\longrightarrow^{\pi^*} \ol\gO^*_0 
\op\longrightarrow^{\pi^1_0{}^*} \ol\gO_1^* \cdots
\op\longrightarrow^{\pi^r_{r-1}{}^*}
 \ol\gO_r^* \longrightarrow\cdots. 
\ee
Its direct limit $\ol\gO^*_\infty$ 
is a presheaf of GDAs on
$J^\infty Y$. Let $\gQ^*_\infty$ be a sheaf constructed from 
$\ol\gO^*_\infty$. The algebra 
$\cQ^*_\infty=\G(\gQ^*_\infty)$ of 
sections of $\gQ^*_\infty$ is a GDA whose
elements
$\f$ possess the following property.
For any point $z\in J^\infty Y$, there exist its open
neighbourhood $U$ and an
exterior form
$\f^{(k)}$ on some jet manifold $J^kY$ such that
$\f|_U= \f^{(k)}\circ \pi^\infty_k|_U$. There is the
monomorphism
$\cO^*_\infty
\to\cQ^*_\infty$ whose image consists of all exterior forms on
finite order jet manifolds.

Restricted to a
coordinate chart (\ref{jet1}), elements of
$\cO^*_\infty$ can be written in a
coordinate form, where horizontal forms 
$\{dx^\la\}$ and contact 1-forms
$\{\th^i_\La=dy^i_\La -y^i_{\la+\La}dx^\la\}$ make up local
generators of the $\cO^0_\infty$-algebra
$\cO^*_\infty$. 
There is the canonical decomposition
\be
\cO^*_\infty =\op\oplus_{k,m}\cO^{k,m}_\infty, \qquad 0\leq k,
\qquad 0\leq m\leq n,
\ee
of $\cO^*_\infty$ into $\cO^0_\infty$-modules $\cO^{k,m}_\infty$
of $k$-contact and $m$-horizontal forms
together with the corresponding
projections $h_k:\cO^*_\infty\to \cO^{k,*}_\infty$ and
$h^m:\cO^*_\infty\to \cO^{*,m}_\infty$.
Accordingly, the
exterior differential on $\cO_\infty^*$ is split
into the sum $d=d_H+d_V$ of the total and vertical
differentials 
\be
&& d_H\circ h_k=h_k\circ d\circ h_k, \qquad d_H\circ
h_0=h_0\circ d, \qquad d_H(\f)= dx^\la\w d_\la(\f), \\ 
&& d_V \circ h^m=h^m\circ d\circ h^m, \qquad
d_V(\f)=\th^i_\La \w \dr^\La_i\f, \qquad \f\in\cO^*_\infty.
\ee
One also introduces the projection $\Bbb R$-module
endomorphism 
\mar{r12}\beq
\vr=\op\sum_{k>0} \frac1k\ol\vr\circ h_k\circ h^n,
\qquad \ol\vr(\f)= \op\sum_{|\La|\geq 0}
(-1)^{\nm\La}\th^i\w [d_\La(\dr^\La_i\rfloor\f)], 
\qquad \f\in \cO^{>0,n}_\infty, \label{r12}
\eeq
of $\cO^*_\infty$ such that
$\vr\circ d_H=0$, and the nilpotent variational operator
$\dl=\vr\circ d$ on $\cO^{*,n}_\infty$. Then, 
$\cO^*_\infty$ is split into the well-known variational
bicomplex. If $Y$ is
contractible, this
bicomplex at terms except $\Bbb R$ is exact. 
This fact is known as the algebraic
Poincar\'e lemma  (e.g., \cite{olv}).

Here, we consider only the variational complex
\mar{b317}\beq
0\to\Bbb R\to \cO^0_\infty
\ar^{d_H}\cO^{0,1}_\infty\cdots  
\op\longrightarrow^{d_H} 
\cO^{0,n}_\infty  \op\longrightarrow^\dl E_1 
\op\longrightarrow^\dl 
E_2 \ar \cdots, \qquad E_k=\vr(\cO^{k,n}_\infty). \label{b317} 
\eeq
One can think of  
\mar{g10'}\beq
L=\cL\om\in \cO^{0,n}_\infty, \qquad \om=dx^1\w\cdots\w dx^n,
\qquad \om_\m=\dr_\m\rfloor\om, \label{g10'}
\eeq
as being
a finite order Lagrangian, while $\dl L$  is its
Euler--Lagrange operator
\mar{g1}\beq
\dl L=\op\sum_{|\La|\geq
0}(-1)^{|\La|}d_\La(\dr^\La_i \cL)\th^i\w\om. \label{g1}
\eeq

\begin{theo} \label{g90} \mar{g90}
Cohomology of the variational complex 
(\ref{b317}) is isomorphic to the de Rham cohomology of the
fiber bundle
$Y$, i.e., $H^{k<n}(d_H)=H^{k<n}(Y)$,
$H^{k-n}(\dl)=H^{k\geq n}(Y)$. 
\end{theo}

\noindent {\it Outline of proof.} \cite{lmp,ijmms} (see
also \cite{and}). We have the complex of
sheaves of
$\cQ^0_\infty$-modules 
\mar{g91}\beq
0\to\Bbb R\to \gQ^0_\infty
\ar^{d_H}\gQ^{0,1}_\infty\cdots  
\op\longrightarrow^{d_H} 
\gQ^{0,n}_\infty  \op\longrightarrow^\dl \gE_1 
\op\longrightarrow^\dl 
\gE_2 \longrightarrow \cdots  \label{g91} 
\eeq
on $J^\infty Y$ and the complex of their structure modules
\mar{g92}\beq
0\to\Bbb R\to \cQ^0_\infty
\ar^{d_H}\cQ^{0,1}_\infty\cdots  
\op\longrightarrow^{d_H} 
\cQ^{0,n}_\infty  \op\longrightarrow^\dl \cE_1 
\op\longrightarrow^\dl 
\cE_2 \longrightarrow \cdots\,.  \label{g92} 
\eeq
Since the paracompact space
$J^\infty Y$ admits a partition of unity by elements of
$\cQ^0_\infty$ \cite{tak2}, the
sheaves of
$\cQ^0_\infty$-modules on
$J^\infty Y$ are acyclic. Then, by virtue of the above
mentioned algebraic Poincar\'e lemma, the 
complex (\ref{g91}) is a resolution of the constant sheaf
$\Bbb R$. In accordance with the abstract de Rham theorem,
cohomology of
the complex (\ref{g92}) equals the cohomology 
of $J^\infty Y$ with coefficients in $\Bbb R$. This
cohomology, in turn, is isomorphic to  the de Rham cohomology
of $Y$, which is a strong deformation retract of $J^\infty Y$
\cite{ander,tak2}. Finally, the $d_H$- and $\dl$-cohomology of
$\cQ^*_\infty$ is proved to equal that of its
subalgebra 
$\cO^*_\infty$ \cite{lmp,ijmms}. \hfill $\Box$ \medskip

A corollary of Theorem \ref{g90} is that 
any $\dl$-closed form $L\in\cO^{0,n}$ is split into the sum
\mar{t42}\beq
 L=h_0\varphi + d_H\xi,  \qquad \xi\in
\cO^{0,n-1}_\infty,
\label{t42}
\eeq
where $\varphi$ is a closed $n$-form on $Y$. In other words, 
a finite order Lagrangian
$L$  is variationally trivial iff
it takes the form (\ref{t42}).

\begin{prop} \label{g93} \mar{g93}
For any Lagrangian $L\in \cO^{0,n}_\infty$, there is the
decomposition 
\mar{+421}\beq
dL=\dl L - d_H(\Xi),
\qquad \Xi\in \cO^{1,n-1}_\infty. \label{+421}
\eeq
\end{prop}

\begin{proof} Let us consider another subcomplex
\mar{xx}\beq
0\to \cO^{1,0}_\infty\ar^{d_H} \cO^{1,1}_\infty
\cdots
\ar^{d_H}\cO^{1,n}_\infty\ar^\vr E_1\to 0
\label{xx}
\eeq
of the variational bicomplex. Similarly to the proof of Theorem
\ref{g90}, one can show that it is exact \cite{lmp,ijmms}. 
 Its exactness at the term
$\cO^{1,n}_\infty$ implies 
the $\Bbb R$-module decomposition
\be
\cO^{1,n}_\infty=E_1\oplus d_H(\cO^{1,n-1}_\infty)
\ee
with  respect to the 
projector $\vr$. Applied to $dL\in\cO^{1,n}_\infty$,
it gives the decomposition (\ref{+421}).
\end{proof}

The form $\Xi$ in the decomposition (\ref{+421}) is not unique.
It reads 
\mar{g43}\beq
\Xi=\op\sum_{s=0}F^{\la\nu_s\ldots\nu_1}_i
\th^i_{\nu_s\ldots\nu_1}\w\om_\la,\qquad 
F_i^{\nu_k\ldots\nu_1}=
\dr_i^{\nu_k\ldots\nu_1}\cL-d_\la F_i^{\la\nu_k\ldots\nu_1}
+h_i^{\nu_k\ldots\nu_1},  \label{g43}
\eeq
where functions $h$ obey the relations $h^\nu_i=0$,
$h_i^{(\nu_k\nu_{k-1})\ldots\nu_1}=0$. It follows that 
$\Xi_L=\Xi+L$ is a Lepagean equivalent, e.g., a
Poincar\'e--Cartan form of a finite order Lagrangian $L$
\cite{got}. The decomposition (\ref{+421}) leads to the
desired first variational formula.

\section{Generalized Lagrangian symmetries}

Let $\gd\cO^0_\infty$ be the Lie algebra of derivations of the 
$\Bbb R$-ring
$\cO^0_\infty$ of smooth real functions of finite jet order on
$J^\infty Y$. 
A derivation $\up\in\gd\cO^0_\infty$ is said to be a
generalized symmetry if the Lie derivative $\bL_\up\f$ of any
contact one-form
$\f\in\cO^{1,0}_\infty$ is also a contact form.
Forthcoming Propositions \ref{g62} -- \ref{g72} confirm
the contentedness of this definition.

\begin{prop} \label{g62} \mar{g62}
The derivation module $\gd\cO^0_\infty$ is isomorphic to the 
$\cO^0_\infty$-dual $(\cO^1_\infty)^*$ of the module of
one-forms $\cO^1_\infty$.
\end{prop}

\begin{proof}
At first, let us show that $\cO^*_\infty$ is generated by
elements $df$,
$f\in \cO^0_\infty$. It suffices to justify that any
element of $\cO^1_\infty$ is a finite $\cO^0_\infty$-linear
combination of elements $df$,
$f\in \cO^0_\infty$. Indeed, every
$\f\in\cO^1_\infty$ is an exterior form on some finite order
jet manifold $J^kY$ and, by
virtue of the Serre--Swan theorem (extended to non-compact
manifolds \cite{ren,ss}), it is represented by a finite sum of
elements $df$, $f\in C^\infty(J^rY)\subset \cO^0_\infty$. 
Any element $\Phi\in (\cO^1_\infty)^*$ yields a derivation
$f\to \Phi(df)$ of the ring $\cO^0_\infty$. Since the module
$\cO^0_\infty$ is generated by elements $df$, $f\in
\cO^0_\infty$, different elements of $(\cO^1_\infty)^*$
provide different derivations of $\cO^0_\infty$, i.e., there
is a monomorphism  $(\cO^1_\infty)^*\to \gd\cO^0_\infty$. By
the same formula, any derivation $\up\in \gd\cO^0_\infty$
sends $df\mapsto \up(f)$ and, since $\cO^0_\infty$ is generated
by elements $df$, it defines a morphism 
$\Phi_\up:\cO^1_\infty\to
\cO^0_\infty$. Moreover, different derivations $\up$ provide
different morphisms $\Phi_\up$. Thus, we have a monomorphism
and, consequently, an isomorphism $\gd\cO^0_\infty\to 
(\cO^1_\infty)^*$. 
\end{proof}

\begin{prop} \label{g60} \mar{g60}
With respect to the atlas
(\ref{jet1}), any derivation $\up\in\gd\cO^0_\infty$ is
given by the coordinate expression
\mar{g3}\beq
\up=\up^\la \dr_\la + \up^i\dr_i +
\op\sum_{|\La|>0}\up^i_\La
\dr^\La_i, \label{g3}  
\eeq
where $\up^\la$, $\up^i$, $\up^i_\La$ are
smooth functions of finite jet order obeying the transformation
law
\mar{g71}\beq
\up'^\la=\frac{\dr x'^\la}{\dr x^\m}\up^\m, \qquad
\up'^i=\frac{\dr y'^i}{\dr y^j}\up^j + \frac{\dr y'^i}{\dr
x^\m}\up^\m, \qquad 
\up'^i_\La=\op\sum_{|\Si|\leq|\La|}\frac{\dr y'^i_\La}{\dr
y^j_\Si}\up^j_\Si +
\frac{\dr y'^i_\La}{\dr x^\m}\up^\m. \label{g71} 
\eeq
\end{prop}

\begin{proof}
Restricted to a coordinate chart (\ref{jet1}), 
$\cO^1_\infty$ is a free $\cO^0_\infty$-module
countably generated by the
exterior forms $dx^\la$, $\th^i_\La$. Then,
$\gd\cO^0_\infty=(\cO^1_\infty)^*$ restricted to this chart 
consists of elements (\ref{g3}), where $\dr_\la$, $\dr^\La_i$
are the duals of $dx^\la$, $\th^i_\La$. The
transformation rule (\ref{g71}) results from the transition
functions (\ref{jet1}). Since the atlas (\ref{jet1}) is finite,
a derivation $\up$ preserves $\cO^*_\infty$.
\end{proof}

The contraction $\up\rfloor\f$ 
and the Lie derivative $\bL_\up\f$,
$\f\in\cO^*_\infty$, obey  
the standard formulas. 

\begin{prop} \label{g72} \mar{g72}
A derivation $\up$ (\ref{g3}) is a generalized
symmetry iff 
\mar{g4}\beq
\up^i_\La=d_\La(\up^i-y^i_\m\up^\m)+y^i_{\m+\La}\up^\m, \qquad
0<|\La|.
\label{g4}
\eeq
\end{prop}

\begin{proof}
The expression (\ref{g4}) results from a direct
computation  similarly to the first part
of the above mentioned B\"acklund theorem. Then, one can 
justify that local functions (\ref{g4}) fulfill the 
transformation law (\ref{g71}).
\end{proof}

Thus, we recover the notion of a generalized symmetry 
in item (iii) in Introduction.   

Any generalized symmetry admits the horizontal
splitting
\mar{g5}\beq
\up=\up_H +\up_V=\up^\la d_\la + (\vt^i\dr_i +
\op\sum_{|\La|>0} d_\La \vt^i\dr_i^\La), \qquad \vt^i=
\up^i-y^i_\m\up^\m,
\label{g5}
\eeq
 relative to the canonical connection
$\nabla=dx^\la\ot d_\la$ on the $C^\infty(X)$-ring
$\cO^0_\infty$ \cite{book00}.
For instance,  let $\tau$ be a vector field on $X$. Then,
the derivation 
$\tau\rfloor (d_H f)$, $f\in \cO_\infty^0$,
is a horizontal generalized symmetry
$\up=\tau^\m d_\m$.
It is easily justified that any vertical generalized symmetry
$\up=\up_V$ obeys the relations
\mar{g6}\beq
\up\rfloor d_H\f=-d_H(\up\rfloor\f),
\qquad \bL_\up(d_H\f)=d_H(\bL_\up\f), \qquad \f\in\cO^*_\infty.
\label{g6}
\eeq 

\begin{prop}  \label{g75} \mar{g75}
Given a  Lagrangian $L\in\cO^{0,n}_\infty$, its 
Lie derivative $\bL_\up L$
along a generalized
symmetry
$\up$ (\ref{g5}) obeys the first variational formula
\mar{g8}\beq
\bL_\up L= \up_V\rfloor\dl L +d_H(h_0(\up\rfloor\Xi_L)) 
+\cL d_V (\up_H\rfloor\om), \label{g8}
\eeq
where $\Xi_L$ is a Poincar\'e--Cartan form of $L$.
\end{prop}

\begin{proof}
The formula (\ref{g8}) comes from the splitting
(\ref{+421}) and the first equality (\ref{g6}):  
\mar{g7}\ben
&& \bL_\up L=\up\rfloor dL + d(\up\rfloor L)
=\up_V\rfloor dL + d_H(\up_H\rfloor L) +\cL d_V
(\up_H\rfloor\om)= \label{g7} \\
&& \qquad  \up_V\rfloor\dl L -\up_V\rfloor
d_H\Xi + d_H(\up_H\rfloor L) +\cL d_V (\up_H\rfloor\om)
= \nonumber \\
&& \qquad \up_V\rfloor\dl L +d_H(\up_V\rfloor\Xi +
\up_H\rfloor L) +\cL d_V (\up_H\rfloor\om), \qquad 
\Xi_L=\Xi+L. \nonumber
\een
\end{proof}

Let $\up$ be a divergence
symmetry of $L$, i.e., $\bL_\up L=d_H\si$, $\si\in
\cO^{0,n-1}_\infty$.
By virtue of the expression (\ref{g7}), this condition implies
that a generalized symmetry $\up$ is projected onto $X$, i.e., 
its components $\up^\la$ depend only on coordinates on $X$. 
Then, the first variational formula (\ref{g8}) takes the form 
\mar{g11}\beq
d_H\si= \up_V\rfloor\dl L +d_H(h_0(\up\rfloor\Xi_L)).
\label{g11}
\eeq
Restricted to Ker$\,\dl L$, it leads to the generalized Noether
conservation law
\mar{g32}\beq
0\ap d_H(h_0(\up\rfloor\Xi_L)-\si). \label{g32}
\eeq
A glance at the expression (\ref{g7}) shows
that a generalized symmetry $\up$ (\ref{g5}) projected onto $X$
is a divergence symmetry of a Lagrangian $L$ iff its vertical
part $\up_V$ is so. Moreover, $\up$ and $\up_V$ lead to
the same conservation law (\ref{g32}). 

Finally, let us obtain the characteristic equation for
divergence symmetries of a Lagrangian $L$. Let a
generalized symmetry $\up$ (\ref{g5}) be projected onto $X$.
Then, the Lie derivative $\bL_\up L$ (\ref{g7}) is a horizontal
density. Let us require that it is a $\dl$-closed form, i.e., 
$\dl(\bL_\up L)=0$.
In accordance with the equality (\ref{t42}), this condition is
fulfilled iff 
\mar{g95}\beq
\bL_\up L=h_0\vf +d_H\si, \label{g95}
\eeq
where $\vf$ is a closed $n$-form on $Y$, i.e., $\up$
is a divergence symmetry of $L$ at least locally. Note
that the topological obstruction $h_0\f$ (\ref{g95}) to
$\up$ to be a global divergence symmetry is at most of first
order. If
$Y\to X$ is an affine bundle, its de Rham cohomology equals
that of
$X$ and, consequently, the topological obstruction
$h_0\vf=\vf$  (\ref{g95}) reduces to a non-exact $n$-form on
$X$.

\section{Lagrangian systems on graded manifolds}

In order to describe Lagrangian systems on a graded manifold,
we start from constructing the corresponding GDA
$\cS^*_\infty$.

Let $(X,\gA_Q)$ be the simple graded manifold constructed from
a vector bundle $Q\to X$.
Its structure ring
$\cA_Q$ consists of sections of the exterior bundle
(\ref{g80}) called graded functions. Given bundle coordinates
$(x^\la,q^a)$ on
$Q$ with transition functions
$q'^a=\rho^a_b q^b$, let
$\{c^a\}$ be the corresponding fiber bases for
$Q^*\to X$, together with the transition functions
$c'^a=\rho^a_bc^b$. Then, $(x^\la, c^a)$ is called the local
basis for the graded manifold $(X,\gA_Q)$ \cite{bart,book00}.
With respect to this basis, graded functions read 
\be
f=\op\sum_{k=0} \frac1{k!}f_{a_1\ldots
a_k}c^{a_1}\cdots c^{a_k}, 
\ee
where $f_{a_1\cdots
a_k}$ are local smooth real functions on $X$.

Let $\gd\cA_Q$ be the Lie
superalgebra of graded derivations
of the $\Bbb R$-ring $\cA_Q$, i.e., 
\be
u(ff')=u(f)f'+(-1)^{[u][f]}fu (f'), 
\qquad f,f'\in \cA_Q,
\ee
where $[.]$ denotes the Grassmann parity. Its elements are
called graded vector fields on $(X,\gA_Q)$. 
 Due to the
canonical splitting
$VQ= Q\times Q$, the vertical tangent bundle 
$VQ\to Q$ of $Q\to X$ can be provided with the fiber bases 
$\{\dr_a\}$, dual of 
$\{c^a\}$. Then,
a graded vector field takes the local form
$u= u^\la\dr_\la + u^a\dr_a$,
where $u^\la, u^a$ are local graded functions, and $u$ acts on
$\cA_Q$ by the rule
\mar{cmp50'}\beq
u(f_{a\ldots b}c^a\cdots c^b)=u^\la\dr_\la(f_{a\ldots b})c^a\cdots c^b +u^d
f_{a\ldots b}\dr_d\rfloor (c^a\cdots c^b). \label{cmp50'}
\eeq
This rule implies the corresponding transformation law 
\be
u'^\la =u^\la, \qquad u'^a=\rho^a_ju^j +
u^\la\dr_\la(\rho^a_j)c^j. 
\ee
Then, one can show that
graded vector fields on a simple graded
manifold are sections of a certain vector bundle
$\cV_Q\to X$ which is locally isomorphic to
$\w Q^*\ot(Q\oplus TX)$ \cite{book00,ijmp}.

Using this fact, one can introduce graded exterior forms on the
graded manifold $(X,\gA_Q)$ as sections of the 
exterior bundle $\op\w^k\cV^*_Q$, where 
$\cV^*_Q\to  X$ is the pointwise $\w Q^*$-dual of $\cV_Q$.
Relative to the dual bases $\{dx^\la\}$ for $T^*X$ and
$\{dc^b\}$ for $Q^*$, graded one-forms read 
\be
\f=\f_\la dx^\la + \f_adc^a,\qquad \f'_a=\rho^{-1}{}_a^b\f_b,
\qquad
\f'_\la=\f_\la +\rho^{-1}{}_a^b\dr_\la(\rho^a_j)\f_bc^j.
\ee
Graded exterior forms constitute the GDA $\cC^*_Q$ with
respect to the graded exterior product $\w$ and the even
exterior differential $d$. Recall the standard formulas
\be
&& \f\w\si =(-1)^{\nm\f\nm\si +[\f][\si]}\si\w \f, \qquad
d(\f\w\si)= d\f\w\si +(-1)^{\nm\f}\f\w d\si, \\
&& u\rfloor(\f\w\si)=(u\rfloor \f)\w\si
+(-1)^{|\f|+[\f][u]}\f\w(u\rfloor\si), \\
&& \bL_u\f=u\rfloor d\f+ d(u\rfloor\f), \qquad
\bL_u(\f\w\si)=\bL_u(\f)\w\si +(-1)^{[u][\f]}\f\w\bL_u(\si).
\ee

Since the jet bundle $J^rQ\to X$ of the vector bundle $Q\to X$
is a vector bundle, let us consider the simple graded manifold
$(X,\gA_{J^rQ})$ constructed from 
$J^rQ\to X$. Its local basis is $\{x^\la,c^a_\La\}$, 
$0\leq |\La|\leq r$,
together with the transition functions
\mar{+471}\beq
c'^a_{\la +\La}=d_\la(\rho^a_j c^j_\La),
\qquad 
d_\la=\dr_\la + \op\sum_{|\La|<r}c^a_{\la+\La}
\dr_a^\La, \label{+471}
\eeq 
where $\dr_a^\La$ are the duals of $c^a_\La$.  
Let $\cC^*_{J^rQ}$ be the GDA of graded
exterior forms on the graded manifold $(X,\gA_{J^rQ})$.
Since $\pi^r_{r-1}:J^rQ \to J^{r-1}Q$ is
a linear bundle morphism over $X$,  it yields the
morphism of graded manifolds 
$(X,\gA_{J^rQ})\to (X,\gA_{J^{r-1}Q})$ and the
monomorphism of the GDAs
$\cC^*_{J^{r-1}Q}\to \cC^*_{J^rQ}$ \cite{book00}. Hence, there
is the direct system of the GDAs 
\be
\cC^*_Q\ar \cC^*_{J^1Q}\ar\cdots 
\cC^*_{J^rQ}\ar\cdots\,.
\ee
Its direct limit $\cC^*_\infty$
consists 
of graded exterior forms on graded manifolds
$(X,\gA_{J^rQ})$,
$0\leq r$, modulo the pull-back identification. 
It is a locally free
$C^\infty(X)$-algebra generated by the elements 
$(1, c^a_\La, dx^\la,\th^a_\La=dc^a_\La -c^a_{\la
+\La}dx^\la)$. 

This construction of odd jets enables one to describe odd and
even variables (e.g., ghosts, ghosts-for-ghosts and
antifields in BRST theory) on the same footing. Let us assume
that a fiber bundle $Y\to X$ in Sections 2--3 is affine, and
let us consider the
$C^\infty(X)$-subalgebra
$\cP^*_\infty$ of the GDA $\cO^*_\infty$ which consists of
exterior forms whose coefficients are polynomial in the fiber
coordinates $y^i_\La$, $|\La|\geq 0$. This property is
coordinate-independent due to the transition functions
(\ref{jet1}). It is readily observed that $\cP^*_\infty$
inherits the structure of a GDA. It is a locally free
$C^\infty(X)$-algebra generated by the elements 
$(1, y^i_\La, dx^\la,\th^i_\La)$. 
Let us consider the $C^\infty(X)$-product of graded
algebras $\cC_\infty^*$ and $\cP^*_\infty$ over their common
subalgebra $\cO^*(X)$. It is a graded algebra
$\cS^*_\infty(Q,Y)$ (or, simply, $\cS^*_\infty$ if there is no
danger of confusion) with respect to the exterior product
$\w$ such that 
\be
f(\psi\w\f)=(f\psi)\w\f=\psi\w(f\f), \qquad
\psi\w\f=(-1)^{|\f||\psi|}\f\w\psi,
\qquad |\psi\w\f|=|\psi|+|\f|
\ee
for all $\psi\in \cC^*_\infty$, $\f\in \cP^*_\infty$ and 
$f\in C^\infty(X)$. Elements of $\cS^*_\infty$ are also
endowed with the Grassmann parity such that $[\f]=0$ for all
$\f\in\cP^*_\infty$. Therefore, we continue to call elements
of the ring
$S^0_\infty$ the graded functions. They are polynomials of
$c^a_\La$ and $y^i_\La$ with coefficients in $C^\infty(X)$. The
sum of exterior differentials on
$\cC_\infty^*$ and $\cP^*_\infty$ makes $\cS^*_\infty$ into a
GDA generated locally by the elements 
$(1, c^a_\La, y^i_\La, dx^\la,\th^a_\La,\th^i_\La)$.  
One can think
of $\cS^*_\infty$ as being the algebra of even and odd variables
on a smooth manifold $X$. In particular, this is the case of the
above mentioned Lagrangian BRST theory on $X=\Bbb R^n$
\cite{barn,bran,bran01}.
Let the
collective symbol
$s^a_\La$ further stand both for its even and odd generating
elements
$c^a_\La$ and $y^i_\La$.

The algebra $\cS^*_\infty$ is decomposed 
into
$\cS^0_\infty$-modules $\cS^{k,r}_\infty$ of
$k$-contact and
$r$-horizontal graded forms.
Accordingly, the graded exterior differential $d$ on 
$\cS^*_\infty$ is split into the sum $d=d_H+d_V$ of 
the total differential $d_H(\f)=dx^\la\w d_\la(\f)$, 
$\f\in \cS^*_\infty$, and the vertical one.
Provided with the projection
endomorphism $\vr$ given by the expression similar to
(\ref{r12}) and the graded variational operator $\dl=\vr\circ
d$, the algebra
$\cS^*_\infty$ is split into the variational
bicomplex. 

Here, we are concerned only with the following three complexes:
\mar{g110-2}\ben
&& 0\to\Bbb R\ar \cS^0_\infty\ar^d \cS^1_\infty\cdots
\ar^d\cS^k_\infty \ar\cdots, \label{g110}\\ 
&& 0\ar \Bbb R\ar
\cS^0_\infty\ar^{d_H}\cS^{0,1}_\infty \cdots
\ar^{d_H} \cS^{0,n}_\infty\ar^\dl 0, 
\label{g111}\\
&& 0\to \cS^{1,0}_\infty\ar^{d_H} \cS^{1,1}_\infty\cdots
\ar^{d_H}\cS^{1,n}_\infty\ar^\vr E_1\to 0, \qquad E_1=
\vr(\cS^{1,n}_\infty). \label{g112}
\een
The first of them is the graded de Rham complex. The second one
is  the short variational complex, where  $L=\cL\om\in
\cS^{0,n}_\infty$ is a graded Lagrangian and 
\mar{g97}\beq
\dl (L)= \op\sum_{|\La|\geq 0}
 (-1)^{|\La|}\th^a\w d_\La (\dr^\La_a L) \label{g97}
\eeq
is its  Euler--Lagrange
operator. The third complex leads us to the first variational
formula.

\begin{theo} \label{g96} \mar{g96} The cohomology of
the complexes (\ref{g110}) -- (\ref{g111}) equals the de
Rham cohomology of 
$X$. The complex (\ref{g112}) is exact.
\end{theo}

\begin{proof}
The proof follows the scheme of the proof of Theorem
\ref{g90}. It is given in Appendix.
\end{proof}

\begin{cor} \label{cmp26} \mar{cmp26}
Every $d_H$-closed form $\f\in\cS^{0,m<n}_\infty$
falls into the sum
$\f=\vf + d_H\xi$, 
where $\vf$ is a closed $m$-form on $X$. Every
$\dl$-closed form $L\in \cS^{0,n}_\infty$ (a variationally
trivial graded Lagrangian) is the sum
$\f=\vf + d_H\xi$, 
where $\vf$ is a non-exact $n$-form on $X$. 
\end{cor}

The exactness of the complex (\ref{g112}) at the term
$\cS^{1,n}_\infty$ results in the following.

\begin{prop} \label{g103} \mar{g103}
Given a graded Lagrangian $L=\cL\om$, there
is the decomposition 
\mar{g99,'}\ben
&& dL=\dl L - d_H(\Xi),
\qquad \Xi\in \cS^{1,n-1}_\infty, \label{g99}\\
&& \Xi=\op\sum_{s=0}
\th^a_{\nu_s\ldots\nu_1}\w
F^{\la\nu_s\ldots\nu_1}_a(\dr_\la\rfloor\om),\qquad 
F_a^{\nu_k\ldots\nu_1}=
\dr_a^{\nu_k\ldots\nu_1}\cL-d_\la F_a^{\la\nu_k\ldots\nu_1}
+h_a^{\nu_k\ldots\nu_1},  \label{g99'}
\een
where graded functions $h$ obey the relations
$h^\nu_a=0$,
$h_a^{(\nu_k\nu_{k-1})\ldots\nu_1}=0$. 
\end{prop}

\begin{proof} The proof repeats that of Proposition
\ref{g93}.
\end{proof}

Proposition \ref{g103} shows the existence of a Lepagean
equivalent 
$\Xi_L=\Xi+L$ of a graded Lagrangian $L$. Locally, one can
always choose $\Xi$ (\ref{g99'}) where all functions $h$ vanish.

\section{Generalized Lagrangian supersymmetries}

Generalized
supersymmetries are defined as graded derivations $\up\in\gd
\cS^0_\infty$ of the $\Bbb R$-ring $\cS^0_\infty$ such that the
Lie derivative $\bL_\up\f$ of any contact graded one-form
$\f\in\cS^{1,0}_\infty$ is also a contact form. Similarly to
the case of generalized symmetries (Propositions \ref{g62} --
\ref{g72}), on can show that any generalized supersymmetry
takes the local form
\mar{g105}\beq
\up=\up_H+\up_V=\up^\la d_\la + (\up^a\dr_a +\op\sum_{|\La|>0}
d_\La\up^a\dr_a^\La), \label{g105}
\eeq
where $\up^\la$, $\up^a$ are local graded functions. Then, it
is easily justified that any vertical generalized supersymmetry
$\up$ (\ref{g105}) obey the relations (\ref{g6}) where $\f\in
\cS^*_\infty$.

\begin{prop}  \label{g106} \mar{g106}
Given a graded Lagrangian $L\in\cS^{0,n}_\infty$, its 
Lie derivative $\bL_\up L$
along a generalized
supersymmetry
$\up$ (\ref{g105}) obeys the first variational formula
\mar{g107}\beq
\bL_\up L= \up_V\rfloor\dl L +d_H(h_0(\up\rfloor \Xi_L)) 
+ d_V (\up_H\rfloor\om)\cL, \label{g107}
\eeq
where $\Xi_L$ is a Lepagean equivalent of $L$.
\end{prop}

The proof is similar to that of Proposition \ref{g75}.
In particular, let $\up$ be a divergence
symmetry of $L$, i.e., $\bL_\up L=d_H\si$, $\si\in
\cS^{0,n-1}_\infty$. 
Then, the first variational formula (\ref{g107}) 
restricted to Ker$\,\dl L$ leads to the 
conservation law
\mar{g108}\beq
0\ap d_H(h_0(\up\rfloor\Xi_L)-\si). \label{g108}
\eeq

The BRST transformation in gauge theory on a principal bundle
$P\to X$ with a structure Lie group $G$ gives an example of a
vertical generalized supersymmetry as follows. Principal
connections on
$P$ are represented by sections of the affine
bundle $C=J^1P/G\to X$ coordinated by $(x^\la, a^r_\la)$
\cite{book,book00,epr}. Infinitesimal generators of
one-parameter groups of vertical automorphism (gauge
transformations) of $P\to X$ are associated to sections of the
vector bundle $V_GP=VP/G$ of right Lie algebras  of the
group $G$. Let us consider the simple graded manifold
$(X,\gA_{V_GY})$ constructed from this vector bundle. Its local
basis is $(x^\la, C^r)$. Let $S^*_\infty(C,V_GP)$ be the above
algebra of even and odd variables $(a^r_\la, C^r)$ on $X$.
Then, the generalized symmetry
\mar{g130}\ben
&& \up= \up_\la^r\frac{\dr}{\dr a_\la^r} +
\up^r\frac{\dr}{C^r}
+\op\sum_{|\La|>0}\left(d_\La\up_\la^r\frac{\dr}{\dr
a_{\La,\la}^r} + d_\La\up^r\frac{\dr}{C^r_\La}\right),
\label{g130}\\ 
&& \up_\la^r=C_\la^r +c^r_{pq}a^p_\la C^q,
\qquad \up^r=
\frac12c^r_{pq}C^p C^q, \nonumber 
\een
is the BRST transformation. The BRST operator is defined as the
Lie derivative $\bs=\bL_\up$ acting on $S^*_\infty(C,V_GP)$.
It is readily observed that it is nilpotent on the
module $S^{0,*}_\infty(C,V_GP)$ of horizontal forms.

Therefore, let us focus on nilpotent generalized
supersymmetries. We say that a vertical generalized
supersymmetry
$\up$ (\ref{g105}) on a GDA $\cS^*_\infty$ is nilpotent if 
\mar{g133}\beq
\bL_\up(\bL_\up\f)= \op\sum_{|\Si|\geq 0,|\La|\geq 0 }
(\up^b_\Si\dr^\Si_b(\up^a_\La)\dr^\La_a + 
(-1)^{[b][\up^a]}\up^b_\Si\up^a_\La\dr^\Si_b \dr^\La_a)\f=0
\label{g133}
\eeq
for any horizontal form $\f\in S^{0,*}_\infty$. A glance at the
second term in the expression (\ref{g133}) shows that a
nilpotent generalized supersymmetry is necessarily odd.
Furthermore, if the equality
\be
\bL_\up(\up^a)=\op\sum_{|\Si|\geq 0} \up^b_\Si\dr^\Si_b(\up^a)=0
\ee
holds for all $\up^a$, a generalized supersymmetry $\up$ is 
nilpotent. A useful example of a nilpotent generalized
supersymmetry is an odd supersymmetry
\mar{g134}\beq
\up=\up^a(x)\dr_a +\op\sum_{|\La|>0}\dr_\La \up^a\dr_a^\La,
\label{g134} 
\eeq
where all $\up^a$ are real smooth functions on $X$, but all
$s^a$ are odd.

Since the Lie derivative $\bL_\up$ and the total
differential $d_H$ mutually commute, let us suppose that the
module of horizontal forms $\cS^{0,*}_\infty$ is split into
a complex of complexes $\{S^{k,m}\}$ with
respect to
$d_H$ and the Lie derivative $\bL_\up$. In order to make
it into a bicomplex, let us introduce the nilpotent operator
$\bs_\up\f=(-1)^{|\f|}\bL_\up\f$, $\f\in S^{0,*}_\infty$,
such that $d_H\circ\bs_\up=-\bs_\up\circ d_H$. This
bicomplex
\be
d_H: S^{k,m}\to S^{k,m+1}, \qquad \bs_\up: S^{k,m}\to
S^{k+1,m} 
\ee
is graded by the form degree $0\leq m\leq n$ and an integer
$k\in\Bbb Z$, though it may happen that $S^{k,*}=0$ starting
from some
$k=k_0$. For short, let us call $k$ the
charge number. 
For instance, the BRST bicomplex
$S^{0,*}_\infty(C,V_GP)$ is graded by the charge number $k$
which is the polynomial degree of its elements 
 in odd variables $C^r_\La$. The bicomplex defined
by the supersymmetry (\ref{g134}) has the similar
gradation, but its nilpotent operator decreases the odd
polynomial degree. 

Let us consider horizontal forms $\f\in \cS^{0,*}_\infty$ such
that a nilpotent generalized supersymmetry $\up$ is their
divergence symmetry, i.e., $\bs_\up\f=d_H\si$. We come to
the relative and iterated cohomology of the
nilpotent operator
$\bs_\up$ with respect to the total differential $d_H$. Recall
that a horizontal form $\f\in S^{*,*}$ is said to be 
a relative $(\bs_\up/d_H)$-closed form if $\bs_\up\f$ is a
$d_H$-exact form. This form is called
exact if it is a sum of an $\bs_\up$-exact form and a
$d_H$-exact form. Accordingly, we have the
relative cohomology $H^{*,*}(\bs_\up/d_H)$. In BRST theory,
it is known as the local BRST cohomology 
\cite{barn,bran}. If a $(\bs_\up/d_H)$-closed
form $\f$ is also $d_H$-closed, it is called an iterated 
$(\bs_\up|d_H)$-closed
form. This form $\f$ is said to be exact if $\f=\bs_\up\xi
+d_H\si$, where $\xi$ is a $d_H$-closed form. Note that the
iterated cohomology $H^{*,*}(\bL_\up|d_H)$ of a
$(\bs_\up,d_H)$-bicomplex
$S^{*,*}$ is exactly the term $E_2^{*,*}$ of its spectral
sequence
\cite{mcl}. There is an obvious isomorphism 
$H^{*,n,}(\bs_\up/d_H)=H^{*,n}(\bs_\up|d_H)$ of relative  and
iterated cohomology groups on horizontal densities.  This
cohomology naturally characterizes Lagrangians
$L$, for which $\up$ is a divergence symmetry, modulo the Lie
derivatives $\bL_\up\xi$, $\xi\in S^{0,*}_\infty$, and the
$d_H$-exact forms. One can apply Theorem 1 in
\cite{lmp} in order to state the relations between the iterated
cohomology and the total $(\bs_\up+ d_H)$-cohomology of the
bicomplex $S^{*,*}$ under the assumptions that all exterior
forms on $X$ are of the same charge number (since
$\up$ is vertical, they are $\bs_\up$-closed) and they are
not $\bs_\up$-exact. This is the case of the BRST
transformation (\ref{g130}), but not the supersymmetry
(\ref{g134}).

\section{Appendix. Proof of Theorem 7}

We start from the exactness of the complexes (\ref{g110}) --
(\ref{g112}), except the terms $\Bbb R$, on $X=\Bbb R^n$. The
Poincar\'e lemma and the algebraic Poincar\'e lemma have been
extended to the complexes (\ref{g110}) and
(\ref{g111}) \cite{barn,bart,bran}.
The algebraic Poincar\'e lemma is applied to  
the complex (\ref{g112}) as follows.

 The fact that a $d_H$-closed graded exterior form
$\f\in \cS^{1,m<n}_\infty$ is $d_H$-exact results from  
the algebraic Poincar\'e
lemma for horizontal graded exterior forms 
$\f\in \cS^{0,m<n}_\infty$.  
Indeed, let us formally associate to an
$(m+1)$-form
$\f=\sum\f_a^\La\w ds^a_\La$ the horizontal
$m$-form $\ol\f=\sum\f_a^\La\ol s^a_\La$ depending on additional
variables $\ol s^a_\La$ of the same Grassmann parity as
$s^a_\La$. It is easily justified that $\ol{d_H\f}=\ol
d_H\ol\f$. If $d_H\f=0$, then $\ol d_H\ol\f=0$ and,
consequently, $\ol \f= \ol d_H \ol\psi$ where
$\ol\psi=\sum\psi_a^\La\ol s^a_\La$ is linear in $\ol
s^a_\La$.   Then, $\f=d_H\psi$ where $\psi=\sum\psi_a^\La\w
ds^a_\La$. It remains to show that, if 
\be
\vr(\f)=\op\sum_{|\La|\geq 0}(-1)^{|\La|}\th^a\w
[d_\La(\dr_a^\La\rfloor\f)]= \op\sum_{|\La|\geq 0}
(-1)^{|\La|}\th^a\w
[d_\La\f_a^\La]=0, \qquad  \f\in
\cS^{1,n}_\infty,
\ee
then $\f$ is $d_H$-exact. A direct
computation gives 
\be
\f=d_H\psi,\qquad  \psi=-\op\sum_{|\La|\geq
0}\op\sum_{\Si+\Xi=\La} (-1)^{|\Si|}\th^a_{\Xi}\w
d_\Si\f^{\La+\m}_a \om_\m.
\ee

Let us associate to each open subset $U\subset X$ the $\Bbb
R$-module $\cS^*_U$ of elements of $\cS^*_\infty$ restricted
to $U$. It is readily observed that these make up a presheaf
on $X$. Let $\gS^*_\infty$ be the sheaf constructed from this
presheaf and $\G(\gS^*_\infty)$ its structure module of
sections. One can show that $\gS^*_\infty$ inherits the 
bicomplex operations, and $\G(\gS^*_\infty)$ does so. 
For short, one can say that $\G(\gS^*_\infty)$ consists of
polynomials in $s^a_\La$, $ds^a_\La$ of
locally bounded jet order $|\La|$. There is the monomorphism
$\cS^*_\infty\to\G(\gS^*_\infty)$. 

Let us consider the
complexes of sheaves of $C^\infty(X)$-modules 
\mar{g113-5}\ben
&& 0\to\Bbb R\ar \gS^0_\infty\ar^d \gS^1_\infty\cdots
\ar^d\gS^k_\infty \ar\cdots, \label{g113}\\ 
&& 0\ar \Bbb R\ar
\gS^0_\infty\ar^{d_H}\gS^{0,1}_\infty \cdots
\ar^{d_H} \gS^{0,n}_\infty\ar^\dl 0, 
\label{g114}\\
&& 0\to \gS^{1,0}_\infty\ar^{d_H} \gS^{1,1}_\infty\cdots
\ar^{d_H}\gS^{1,n}_\infty\ar^\vr \gE_1\to 0,
\label{g115}
\een
on $X$ and the complexes of their structure modules
\mar{g116-8}\ben
&& 0\to\Bbb R\ar \G(\gS^0_\infty)\ar^d \G(\gS^1_\infty)\cdots
\ar^d\G(\gS^k_\infty) \ar\cdots, \label{g116}\\ 
&& 0\ar \Bbb R\ar
\G(\gS^0_\infty)\ar^{d_H}\G(\gS^{0,1}_\infty) \cdots
\ar^{d_H} \G(\gS^{0,n}_\infty)\ar^\dl 0, 
\label{g117}\\
&& 0\to \G(\gS^{1,0}_\infty)\ar^{d_H} \G(\gS^{1,1}_\infty)\cdots
\ar^{d_H}\G(\gS^{1,n}_\infty)\ar^\vr \G(\gE_1)\to 0.
\label{g118}
\een
The complexes (\ref{g113}) -- (\ref{g114}) are resolutions
of the constant sheaf $\Bbb R$, while the complex (\ref{g115})
is exact.  By virtue of the abstract de Rham theorem, the
cohomology of the complexes (\ref{g116}) --
(\ref{g117}) equals the de Rham cohomology $H^*(X)$ of $X$,
whereas the complex (\ref{g118}) is exact. It remains to prove
that cohomology of the complexes (\ref{g110}) -- (\ref{g112})
equals that of the complexes (\ref{g116}) -- (\ref{g118}).
The proof follows that of Theorem 9 in \cite{lmp} and Theorem
5.1 in \cite{ijmms}.

Let the common symbols $\G^*_\infty$ and $D$ stand for all
modules and operators in the complexes (\ref{g116}) --
(\ref{g118}), respectively. With this notation, one can say that
any
$D$-closed element
$\f\in\G^*_\infty$ takes the form $\f=\vf + D\xi$, where $\vf$
is an exterior form on $X$. Then, it suffices to show
that, if an element
$\f\in
\cS^*_\infty$ is
$D$-exact in the module $\G^*_\infty$, then it is
so in $\cS^*_\infty$.
By virtue of the above mentioned Poincar\'e lemmas, 
if $X$ is contractible and a $D$-exact element $\f$ 
is of finite jet order
$[\f]$ (i.e., $\f\in\cS^*_\infty$), there exists an 
element $\vf\in
\cS^*_\infty$ such that $\f=D\vf$. Moreover,
a glance at the corresponding homotopy operators shows that 
the jet order
$[\vf]$ of $\vf$ is bounded by an integer $N([\f])$,
depending only on $[\f]$. We agree to call this fact the finite
exactness of the operator
$D$. Given an arbitrary manifold
$X$, the finite exactness takes place on 
any domain $U\subset X$. The following statements are proved
similarly to those in \cite{lmp,ijmms}. 

(i) Given a family $\{U_\al\}$ of disjoint open subsets of $X$,
let us suppose that the finite exactness takes place on
every subset $U_\al$. Then, it is true on the union
$\op\cup_\al U_\al$.

(ii) Suppose that the finite exactness of the operator $D$ takes
place on open subsets
$U$, $V$ of $X$ and their non-empty overlap $U\cap V$. Then, it
is also true on $U\cup V$.

It remains to choose an appropriate cover of $X$. It admits a
countable cover $\{U_\xi\}$ by domains $U_\xi$, $\xi\in \Bbb
N$, and its refinement $\{U_{ij}\}$, where $j\in\Bbb N$ and $i$
runs through a finite set, such that $U_{ij}\cap
U_{ik}=\emptyset$, $j\neq k$ \cite{greub}. Then, $X$ has the
finite cover $\{U_i=\bigcup_j U_{ij}\}$. Since the finite
exactness of $D$ takes place on any domain $U_\xi$, it also 
holds on any member $U_{ij}$ of the refinement $\{U_{ij}\}$ of
$\{U_\xi\}$ and, in accordance with the assertion (i),
on any member of the finite cover $\{U_i\}$ of $X$. Then, the
assertion (ii) states the finite exactness of $D$  on
$X$.

\end{document}